\documentclass[a4paper]{jpconf}
\usepackage{graphicx}
\usepackage{url}
\begin{document}
\title{Econoinformatics meets Data-Centric Social Sciences}

\author{Aki-Hiro Sato}

\address{Department of Applied Mathematics and Physics, Graduate School
of Informatics, Kyoto University, Yoshida-Honmachi, Sakyo-ku 606-8501,
Kyoto JAPAN}

\ead{sato.akihiro.5m@kyoto-u.ac.jp}

\begin{abstract}
Our society has been computerised and globalised due to emergence and spread
of information and communication technology (ICT). This enables us to
investigate our own socio-economic systems based on large 
amounts of data on human activities. In this article, methods of treating
complexity arising from a vast amount of data, and linking data from
different sources, are discussed. Furthermore, several examples are
given of studies into the applications of econoinformatics for the
Japanese stock exchange, foreign exchange markets, domestic hotel booking
data and international flight booking data are shown. It is the main
 message that spatio-temporal information is a key element to synthesise
 data from different data sources.
\end{abstract}

\section{Introduction}
Recently people feel data in various fields due to 
developments in information and communication technology
(ICT). This circumstance enables us to collect, store,
and analyse data on our socio-economic environment in order to
interpret our society and to pursue evidence-based management. One can collect
and accumulate large amounts of socio-economic data on human
activities, and analyse and visualise them. Vast
  amounts of data collected from socio-economic systems have made
  new types of commercial services  
and research fields emerging. According to
M.F. Goodchild~\cite{Goodchild}, every human 
is able to act as an intelligent sensor, and in that sense, the earth's
surface is currently occupied by more than six billion sensors. 

We can extract information from an amount of data, construct knowledge
from lots of information, and hopefully establish wisdom from several
pieces of knowledge. Specifically, researchers in the fields of sociology,
economics, informatics, and physics are focusing on these frontiers and
have launched data-centric social sciences in order to understand
the complexity of socio-economic systems~\cite{futurict}. 

However, since our society which is the sum total of both internal and 
external states of individuals is several orders of magnitude more
complicated than each individual, it seems difficult to
image how we could manage to capture the real totality of 
its states with the cooperation of many agencies. The nature of
this problem is referred to recognised as ``complexity'', which is a new
research field into knowledge on groups of people,
organisations, communities, and the economy actually behave in the real
world.

Information theory describes the number of possible states of a system
as its ``complexity''. Complexity may be estimated with
Shannon information entropy or thermodynamic 
entropy. For example, researchers have used a methodology to characterise
the structure of networks with information-theoretic
entropy~\cite{Dehmer:11,Wilhelm}. According to a study by Dehmer
and Mowshowitz~\cite{Dehmer:11}, the concept of graph entropy was first 
proposed in the 1950s to measure structural
complexity. Rashevsky~\cite{Rashevsky:55}, Trucco~\cite{Trucco:56}, and
Mowshowitz~\cite{Mowshowitz:68} were the first researchers to define and
investigate the entropy of graphs. Several graph invariants, such as the
number of vertices, vertex degree sequence, and extended degree
sequence, have been used in the construction of entropy-based
measures. Wilhelm an Hollunder considered the
normalised weight of the flux between two nodes as the probability 
of a symbol in the transmitter signal that corresponds to the sum of all
influxes to/effluxes from a given node~\cite{Wilhelm}. Sato also considered
information-theoretic measures for a bipartite graph, and inferred
economic situations  using the network entropy of relative frequencies
among group populations~\cite{Aki}. Bianconi looks at the entropy measure
of network ensembles under several constraints~\cite{Bianconi,Anand}.

According to Heinz von Foerster~\cite{Foerster}, complexity is not 
a property which observed systems possess rather, it is to be perceived
by observing systems. He asks us about it through the following
question: \textit{Are the states of order and disorder states of affairs
that have been discovered, or are their states of affairs that are
invented?} If states of order and disorder are discovered, then
complexity is a property of the observed systems. If invented then
it pertains to the observing systems. Foerster's definition of
complexity proposes that the relative degree of order and disorder is
determined by the degrees of freedom of an observed system and an
observing system. One of the most significant reasons that we recognise
complexity in observed systems is because of the finiteness of periods when 
and abilities with which we are able to observe the systems, and
the limitations of our memory and a priori knowledge of them that lead
to our bounded rationality, or nescience.

From the data-centric perspective, complexity is
referred to as a ratio of an amount of data to human cognitive
capacity. How many megabytes of memory the human brain has? Von
Neumann proposed that human memory is estimated as 100 exabits from 
all the neural impulses conducted in the brain during a
lifetime~\cite{von-Neumann}. Another method is to estimate the total
number of synapses, and then presume that each synapse can hold a few
bits. Estimates of the number of synapses have been made in the range
from $10^{13}$ to $10^{15}$, with corresponding estimates of memory
capacity. Landauer investigated how much people remember at Bell Communications
Research~\cite{Landauer}. The remarkable result of this work was that
human beings remembered very nearly two bits per second under all the
experimental conditions (visual, verbal, musical). Therefore, the 35-year
accumulation of human beings' memory estiamtes of 0.2 to 1.4 gigabits
if the loss of memory are assumed.

If the amount of data is greater than this estimates
of memory, then all the data can not be memorised by each human
being. Therefore, the amounts of data may determine the
complexity in the sense of the data accessiblity. This may provide a kind of
definition of big-data in data-centric sciences.  

Clearly, we also need to carefully consider methods of
collecting, storing, handling, and analysing a vast amount of data
in computer systems.

In this article, I present exemplar studies of observation and data
analysis with a large amount of data in several socio-economic systems. I
would like to call such studies ``econoinformatics''. Econoinformatics
needs powerful computer systems to visualise and quantify the behaviour of
human beings, adapt to their changing environments, and evolve over time.

Furthermore, we should mention the problem of data linkage in
data-centric studies. Data linkage is referred as to a method of linking
data from different sources with the same elements. If we can synthesise
data from different sources, we may find new insights from the
synthesised data. The comprehensive study is further useful to obtain 
new findings on our environment.

The literature review on data-centric social sciences is provided
in Section \ref{sec:literature} of this article. In
Section \ref{sec:exemplar} exemplar studies on econoinformatics are
shown. The Japanese stock exchange, the foreign exchange market,
hotel booking data, and flight booking data are analysed. In Section
\ref{sec:discussion}, a method of synthesising data from different sources,
and a method of econoinformatics, are discussed. Section
\ref{sec:conclusion} is devoted to conclusions.

\section{Literature review}
\label{sec:literature}

Recently, several researchers in a wide spectrum of fields have paid a
remarkable amount of attention to massive amounts of comprehensive
data. For example, search engines of web services need massive data
about hyperlink connections among web pages, and electronic commerce
systems need to cover information of various kinds of products. Due
to the development of ICT, the Advanced Information Society has already
emerged globally and it has gradually made our world smaller and
smaller. The term ``information explosion'' has been coined to
describe this situation and has been
realised~\cite{infoplosion,Korth}. This term refers to a situation in
which the total amount of information created by individuals exceeds the
individuals' information processing capability.

Studies with vast amounts of socio-economic data have
several branches. Here, five kinds of recent studies (financial market
data, demographic data, traffic flow data, POS data, and e-commerce
data) are surveyed for the purpose of finding ways of coping with the
complexity of human societies.

A large amount of data on financial markets is available because the
electronic matching systems of financial markets are spreading all over
the world due to the development of ICT. Contemporary trading is done through
electronic platforms, and settlement operations are done through electronic
clearing systems. Financial market data can be collected through a
direct application programming interface (API) or through the historical
data centres of data providers. Applications of statistical mechanics to
finance, by means of statistical physics, agent-based modelling, and
network analysis, have progressed during the last decade~\cite{Takayasu,
Mantegna, Sornette, Aki:08}.

The launch of the E-Stat database by the Japanese
government~\cite{estat} provides us 
with new technological means for a data-based understanding of our
country. In principle, everyone can understand the state of our country
from demographic data. Furthermore,
real-time demographic data are also available since the technologies
to collect human activities via personal mobile phones have been
established~\cite{Gonzalez}. In the near future, 
we will be able to visualise real-time demographics, both comprehensively
and circumstantially. 

Recently, several car navigation companies have launched autonomous
sensory navigation services in Japan. As a result, these companies can
collect real-time car traffic data via each car navigation
terminal. By collecting data from many cars, one can find roads
and points where traffic jams are occurring. Without constructing new
infrastructure to collect traffic states, real-time traffic data can be
accumulated due to the development of Integrated Transport Systems 
(ITS). Based on such data, comprehensive analyses of traffic flows can be
conducted in order to cope with traffic jams~\cite{Antoniou}. Recent
developments in traffic measurement technologies have been driving the theoretical
development of traffic control and modelling~\cite{Helbing}. 

POS is an abbreviation for ``point-of-sales'', and all department
stores and supermarkets have introduced this kind of system in order to
ring up purchases at cash registers. As a result, retail sales can
be managed in real-time, and data-centric operations can be done. On the
basis of these massive amounts of data, new marketing methods have been
developed. The statistical properties of expenditure in a single
shopping trip show a power-law distribution~\cite{Mizuno}. A comprehensive
analysis of retail sales is one of the most promising directions to be
followed in order to bridge the gap between microeconomics and macroeconomics. 

Web-based commerce systems enable us to purchase everything, from books
to electronic equipment, via websites. The details of consumers and
goods can be stored on the data-base engine of each website. If we can
use such data, then we may, in principle, capture real-time demand and
supply of all items which are traded via websites. Analysing
massive amounts of data on items which are sold via web commerce systems
is expected to open a window to new economic theory and service
engineering~\cite{Lambiotte,Deshatres}. Data on hotel booking
opportunities~\cite{Aki:02}, international flight booking
opportunities~\cite{Aki:02b}, and price comparison sites~\cite{Mizuno:10} 
has also been studied.

The property that these studies seem to have in common is their 
ability to overcome the complexity in socio-economic systems, by using
massive amounts of data and vast computations. Copious amounts of data
on human activities are collected 
by means of ICT, and vast amounts of computation for such data are
conducted for the purposes of searching, matching, visualising, and
extracting. 

\section{Exemplar studies on econoinformatics}
\label{sec:exemplar}

Large-scale data on socio-economic systems may make it possible to
understand our society from a holistic point of view. To do so, we need
to discuss the possibilities of comprehensive analysis and 
data linkage. In this section, four types of exemplar
studies on socio-economic systems are shown. These examples include
studies on the Japanese stock exchange, the foreign exchange market,
hotel bookings through an e-commerce platform, flight bookings 
through an e-commerce platform. In order to archive 
a comprehensive point of view, both the Japanese and international
economic situations are measured.

\subsection{Japanese stock exchange}
During the last two decades, it is said that the main problem faced by
Japan is a low productivity growth rate. Despite these circumstances, 
the Japanese economy recovered eventually, with the aid of the global growth
from 2004 to 2007. However, the latest global financial crisis strongly
affected many countries, including Japan, after the midquarter of 2008. 
At that time, we observed that Japanese stock prices dropped
sharply due to financial turmoil of 2008-2009. Furthermore, 
huge earthquakes hit Japan on 11 March 2011, and the Japanese economy 
suffered from a large number of social losses.

So far, we have explained the affairs by the Japan explanatory
narrative. The question remains, however, as to how we can 
describe our macroeconomics situations in a more 
quantitative fashion.

Macroeconomic situations strongly influence money flows at all levels
of society. Stocks at each sector are traded by investors and 
traders every minute through the stock exchange market. Moreover, stock
prices are so sensitive to money flows, that stock prices in all
sectors reflect demand-supply gaps of the money by economic
actors. Therefore, they are expected to be useful for detecting
changes in macroeconomic situations.

In this study, we hope to provide some insights on the problem of the
quantification of Japanese macroeconomic situations, through a
comprehensive analysis of stock prices traded in the Tokyo Stock
Exchange. In the context of economics and finance, there are various
methods available for segmenting highly nonstationary financial 
time series into stationary segments, called regimes or
trends. Following the pioneering works of Goldfeld and
Quandt~\cite{Goldfeld}, there is an enormous body of literature on
detecting structural breaks or change points separating stationary
segments. Recently, a recursive entropic scheme to segment financial
time series was proposed~\cite{Siew:11}.

In this study, a recursive segmentation procedure is applied to
an analysis of security prices of 1,413 Japanese firms listed on the first
section of the Tokyo Stock Exchange. The number of segments in quintiles 
in terms of variance is computed in order to detect change points of
money flows of the Japanese security market.

Let $O_{i,t}$ and $E_{i,t}$ be daily opening and closing prices of $i$-th
stock $(i=1,\ldots,M)$ at day $t \quad (t=1,\ldots,N)$. $M$ and $N$ are
denoted as the total number of stock and the observation length.
The daily log-return (opening to closing) time series $x_{i,t} \quad 
(i=1,\ldots,M; t=1,\ldots,n)$ is computed as 
$x_{i,t} = \log E_{i,t} - \log O_{i,t}$ and $n=N-1$. 

According to the seminal work by Mantegna and Stanley~\cite{Mantegna},
the log-return time series of stock prices are modelled by L\'evy
distributions. Superstatistics suggested that a mixture of Gaussian
distributions with $\chi^2$-distributions in terms of variance gives
a L\'evy distribution. Therefore, we assume that each segment is
sampled from a Gaussian distribution with different mean and
variance. It is assumed that the log-return time series in
segment $m_i$ follows a stationary Gaussian distribution with mean
$\mu_{i,m_i}$ and variance $\sigma_{i,m_i}^2$.

To find the unknown segment boundaries $t_{i,m_i}$ separating
segment $m_i$ and $m_i+1$, the recursive segmentation scheme introduced
by Bernaola-Galv\'an et al~\cite{Siew:11,Bernaola} is employed. In this
segmentation scheme, the likelihood of the total time series is compared
with the likelihood of the two segmented time series.

Suppose that there are $n$ observations $x_s \quad (s=1,\ldots,n)$. 
Let $g(x;\mu,\sigma)$ be a Gaussian distribution:
\begin{equation}
g(x;\mu,\sigma^2) =
 \frac{1}{\sqrt{2\pi\sigma^2}}\exp\Bigl[-\frac{(x-\mu)^2}{2\sigma^2}\Bigr],
\label{eq:gauss}
\end{equation}
Assuming that the observations $x_s$ should be segmented at $t$, and that
the observations on the left hand side are sampled from
$g(x;\mu_L,\sigma_L^2)$, and that those on the right hand side are from
$g(x;\mu_R,\sigma_R^2)$, we define likelihood functions:
\begin{eqnarray}
L &=&
 \prod_{s=1}^{n}g(x_s;\mu,\sigma^2), \\
\nonumber
L_2(t) &=& 
 \prod_{s=1}^{t}g(x_s;\mu_L,\sigma_L^2)\prod_{s=t+1}^{n}g(x_s;\mu_R,\sigma_R^2). \\
\end{eqnarray}
Furthermore, we define the logarithmic difference between $L$ and $L_2(t)$ as,
\begin{equation}
\Delta(t) = \log L_2(t) - \log L.
\label{eq:def-delta}
\end{equation}
Inserting Eq. (\ref{eq:gauss}) into Eq. (\ref{eq:def-delta}), we have,
\begin{equation}
\Delta(t) = \sum_{s=1}^t\log
 g(x_s;\mu_L,\sigma^2_L) + \sum_{s=t+1}^n\log g(x_s;\mu_R,\sigma^2_R) - 
\sum_{s=1}^n\log g(x_s;\mu,\sigma^2).
\end{equation}

In general, if a random varialbe $A$ is given and its
  distribution admits a probability density function $f$, then the
  expected value of $A$ (if exists) can be calculated as 
\begin{equation}
\mbox{E}[A] = \int_{-\infty}^{\infty}a f(a)\mbox{d}a,
\end{equation}
where $f(a)$ is a probability density function of $A$. This implies
that the arithmetic mean of $m$ random variables $a_i \quad
(i=1,\ldots,m)$ is approximated as
\begin{equation}
\frac{1}{m}\sum_{i=1}^m a_i \approx \int_{-\infty}^{\infty}a f(a)\mbox{d}a,
\end{equation}
for a sufficiently large value of $m$.

By using this approximation, for $m$ observations
  $x_s \quad (s=1,\ldots,m)$, we obtain
\begin{equation}
\sum_{s=1}^m \log g(x_s;\mu,\sigma^2) \approx
m \int_{-\infty}^{\infty} g(x;\mu,\sigma^2) \log g(x;\mu,\sigma^2)
 \mbox{d}x = -\frac{m}{2}\log \Bigl(2\pi e \sigma^2\Bigr),
\end{equation}
and $\Delta(t)$ is rewritten as,
\begin{equation}
\Delta(t) = n \log \sigma - t \log \sigma_L - (n-t)\log \sigma_R \geq 0.
\label{eq:delta1}
\end{equation}

$\Delta(t)$ can be used as an indicator to separate the observations
into two parts. An adequate way to separate the observations is
that a segmentation is conducted at $t$ where $\Delta(t)$ takes the
maximum value. Namely, an adequate segmentation should be done at,
\begin{equation}
t^* = \arg \max \Delta(t).
\end{equation}
If $\max \Delta(t)$ is less than a threshold value $\Delta_c$, then the
segmentation should be terminated. The hierarchical segmentation
procedure is also applied to the time series. After segmentation,
we also apply this procedure for each segment recursively. In order to 
stop the segmentation procedure we assume that the minimum value
$\Delta_c$. If $\max \Delta(t)$ is less than $\Delta_c$, then we do not
apply the segmentation procedure any more. This is used as the stopping
condition of the recursive segmentation procedure.

\begin{figure}[h]
\centering
\includegraphics[scale=0.9]{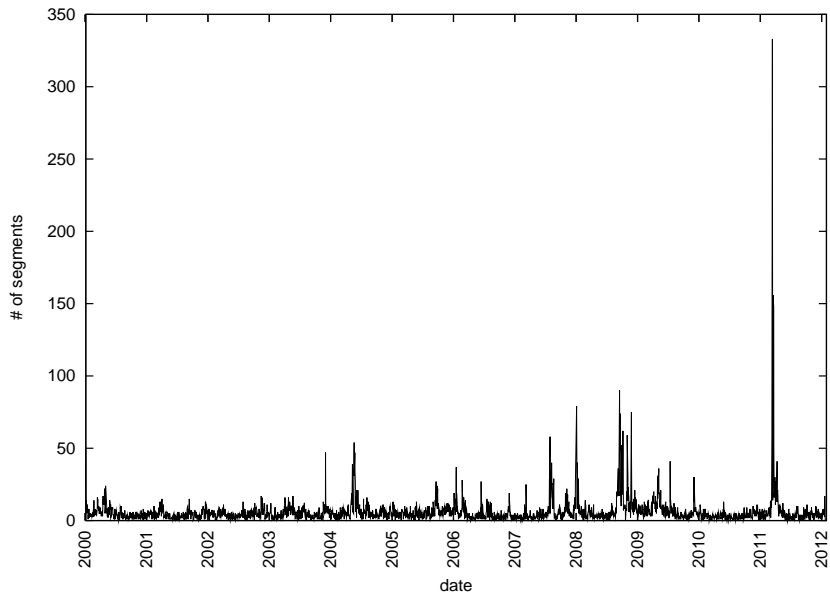}
\caption{The number of starting dates of segments for 1,413 log-return
 time series during 4th January 2000 to 30th January 2012.}
\label{fig:count}
\end{figure}

1,413 companies listed on the first section of Tokyo Stock
Exchange are selected for empirical analysis. The duration is 4 January,
2000 to 30 January, 2012. These companies maintain during the observation
period. The recursive segmentation procedure was applied to
1,413 security prices. The segmentation analysis of daily log-return
time series for ending prices was conducted. Throughout the
investigation $\Delta_c$ is fixed as 
10. Fig. \ref{fig:count} shows the number of starting dates of segments
for 1,413 log-return time series. The number of segments increase
at June 2001, April 2004, February 2006, and 2007 to 2009. These seem to
correspond to regimes or change points in the Japanese
economy. Specifically, during the last global financial crisis the
number of segments tends to increase (about 260 segments can be found at
this period). Furthermore, after 11 March, 2011, the Great East Japan
Earthquake, the number of segments steeply increased larger than during
the last global financial crisis.

\begin{figure}[h]
\centering
\includegraphics[scale=0.9]{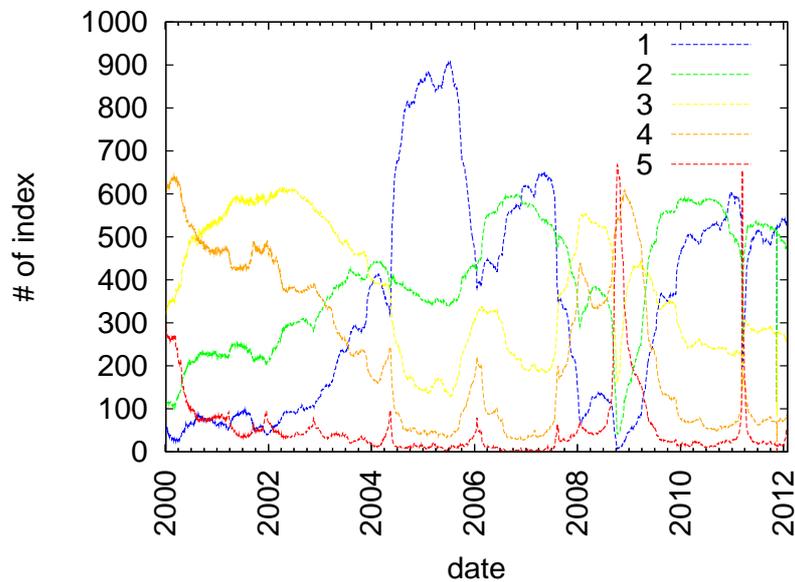}
\caption{The number of segments in each quintile for 1,413 log-return
 time series between 4th January 2000 to 30th January 2012.}
\label{fig:quintile2}
\end{figure}

The number of segments belonging to the same quintile of variances is
counted. Compute order statistics of variance $\{\sigma_{i,(1)} \leq
\ldots \leq \sigma_{i,(m_i)}\}$ in all the segments of stock $i$. 
Next, each segment of stock $i$ is labelled $k=\{1,2,3,4,5\}$,
depending on variance which belongs to the quintile. The number of
segments which have the same labels is counted on each
day. Fig. \ref{fig:quintile2} shows the number of segments belonging to
each quintile on every day. The number of first quintile segments 
shows stability of economic affairs, and the number of fifth quintile
segments indicates instability of economic affairs. It is found that
from 2003 to 2007, the Japanese economy was in a stable regime. From the
end of 2007 an unstable regime was observed. Specifically, during September
2008, when we experienced the Lehman shock, the number of fifth quintile
regimes steeply increased. This implies that the money flows of the Japanese
economy became unstable just after the Lehman shock. From March 2009, the
money flow eventually recovered, and the number of unstable regimes
decreased while the number of stable segments eventually increased. From
11 March to 10 April, 2011 the number of unstable regimes steeply
increased due to the Great East Japan Earthquake. However, the number of
stable regimes rapidly increased thereafter, and Japanese 
macroeconomic affairs have been eventually recovered. 

\subsection{Foreign exchange market}
The foreign exchange market is the largest financial market in the
world. The foreign exchange market is organised by international banks
and funds through brokerage platforms. Since the currency exchange strongly
reflects and is related to international economies, we can predict global
economic conditions from the foreign exchange market. 

\begin{figure}[h]
\centering
\includegraphics[scale=0.6]{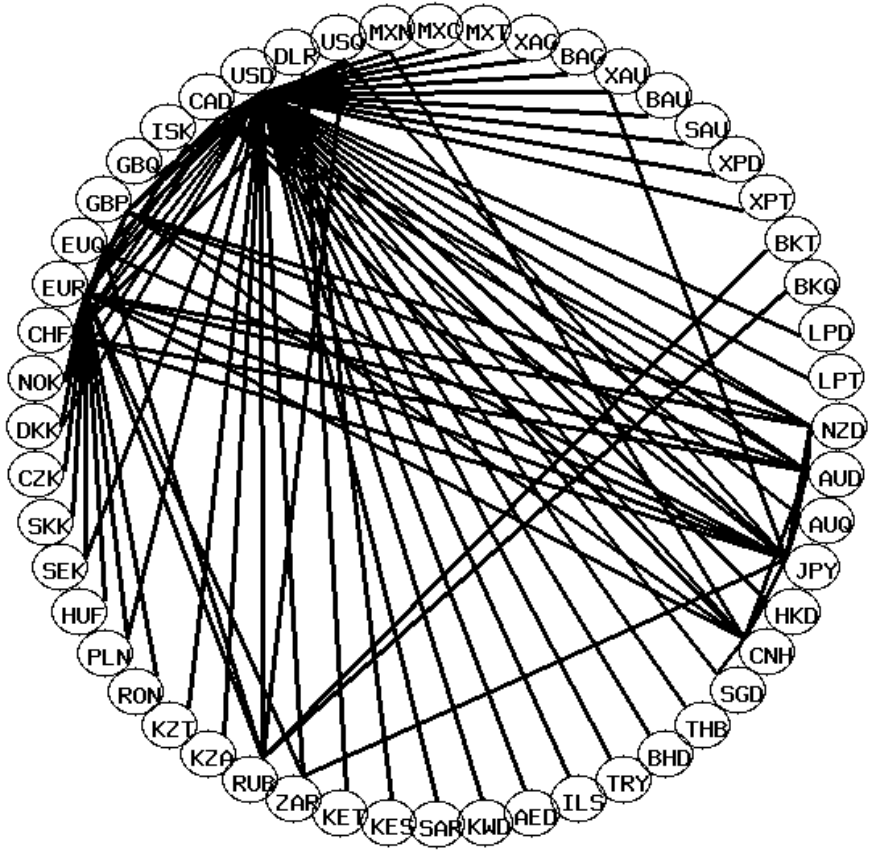}
\caption{Network representation of currencies and currency pairs. Nodes
 represent currencies, and links between two nodes represent currency
 pairs included in the data set.}
\label{fig:network}
\end{figure}

The analysis is conducted using high-resolution data (ICAP EBS Data
Mine Level 1.0) collected using the ICAP EBS platform. The data includes
currency pairs, quotation prices, and transaction prices with a one-second
resolution. In the exchangeable currency pairs, consisting of 41
currencies and 11 precious metals, 93 kinds of currency pairs are
included during the period from June 2008 to June 2012. There are about
520 million records in the observation period. 
Fig. \ref{fig:network} shows exchangeable currency pairs. USD
and EUR have a lot of links to other currencies. AUD, CNH, CHF, JPY, and
GBP have five to eight links. Others have only one or two links to major
currencies.

Suppose one can observe quotations/transactions about the $i$-th
currency, and the $j$-th currency and count the arrival of quotations
or occurrence of transactions for each currency pair on a brokerage
system with an interval of $\Delta (>0)$. This activity is defined as
the number of quotations/transactions which market participants enter
into the electronic brokerage system per $\Delta$. We define
$f_{ij}(t;S)$ as the quotation activities between the $i$-th currency
and $j$-th currency ($i,j = 1,2,\ldots,N$) in $[t\Delta,(t+1)\Delta]
\quad (t=1,2,\ldots,T)$  on the $S$-th observation period. In this
analysis, we adopt the definition that the activities should be counted
in symmetrical manner $f_{ij}(t;S)=f_{ij}(t;S)$ and satisfy no self-dealing
condition $f_{ii}(t;S)=0$. Then, the density of quotations between the
$i$-th currency and $j$-th currency can be estimated as,
\begin{equation}
A_{ij}(S)= \frac{\sum_{t=1}^T
 f_{ij}(t;S)}{\sum_{t=1}^T\sum_{i=1}^N\sum_{j=1}^N f_{ij}(t;S)}.
\end{equation}
Obviously, it has probabilistic properties, such that,
$\sum_{i=1}^N\sum_{j=1}^N A_{ij}(S)=1$, $A_{ij}(S)=0$, and $0 \leq
A_{ij}(S) \leq 1$.

Under the assumption that the attention of market participants to the
exchangeable currency pairs can be estimated as the centrality of
currency pairs, $A_{ij}(S)$ can be empirically estimated by using
quotation/transaction frequencies calculated from high-resolution data
without knowledge of the network structure of market participants. Moreover,
relative occurrence rates of the $i$-th currency on the $S$-th
observation period are defined as,
\begin{equation}
K_i(S) = \sum_{j=1}^N A_{ij}(S),
\end{equation}
where it also has probabilistic properties, such that, 
$\sum_{i=1}^N K_i(S) = 1$ and $0 \leq K_i(S) \leq 1$. Since both $A_{ij}(S)$ and
$K_i(S)$ may be regarded as fingerprints representing the market states
on the observation period $S$, their shape may describe market states at
$S$. To capture the difference of shapes between $S_1$ and $S_2$,
Jensen-Shannon divergence~\cite{Lin:91} is employed. 
\begin{eqnarray}
D_{cp}(S_1,S_2) &=& H_A\Bigl(\bigl\{\frac{1}{2}\sum_{k=1}^2 A_{ij}(S_k)\bigr\}\Bigr)  -
 \frac{1}{2}\sum_{k=1}^2 H_A\Bigl(\bigl\{A_{ij}(S_k)\bigr\}\Bigr) \\
D_{c}(S_1,S_2) &=& H_K\Bigl(\bigl\{\frac{1}{2}\sum_{k=1}^2 K_{i}(S_k)\bigr\}\Bigr)  -
 \frac{1}{2}\sum_{k=1}^2 H_A\Bigl(\bigl\{K_{i}(S_k)\bigr\}\Bigr), 
\end{eqnarray}
where $H_A(A_{ij})$ and $H_K(K_i)$ are, respectively, denoted as the
Shannon entropies defined as,
\begin{eqnarray}
H_A(\{A_{ij}\}) &=& -\sum_{i=1}^N\sum_{j=1}^N A_{ij}\log A_{ij}, \\
H_K(\{K_i\}) &=& -\sum_{i=1}^N K_i \log K_i.
\end{eqnarray}
From these definitions, we can confirm that they have the following
properties:
\begin{eqnarray}
D_{cp}(S_1,S_2) &=& D_{cp}(S_2,S_1), \\
D_{cp}(S_1,S_2) &\geq& 0, \\
D_{cp}(S_1,S_2) &=& 0 \quad \mbox{iff} \quad A_{ij}(S_1) = A_{ij}(S_2)
 \forall{i,j}, \\
D_{c}(S_1,S_2) &=& D_{c}(S_2,S_1), \\
D_{c}(S_1,S_2) &\geq& 0, \\
D_{c}(S_1,S_2) &=& 0 \quad \mbox{iff} \quad K_{i}(S_1) = K_{i}(S_2)
 \forall{i}.
\end{eqnarray}

The similarity of market states between two observation periods is
computed for each week. The Jensen-Shannon divergence is employed in
order to compute similarities 
since it has resistance characteristic for zero probabilities. This
means that the Shannon entropy does not diverge to infinity because of 
$0 \log 0 = 0$, which is proven in \ref{sec:zero-log-zero}.

\begin{figure}[hbt]
\centering
\includegraphics[scale=0.34]{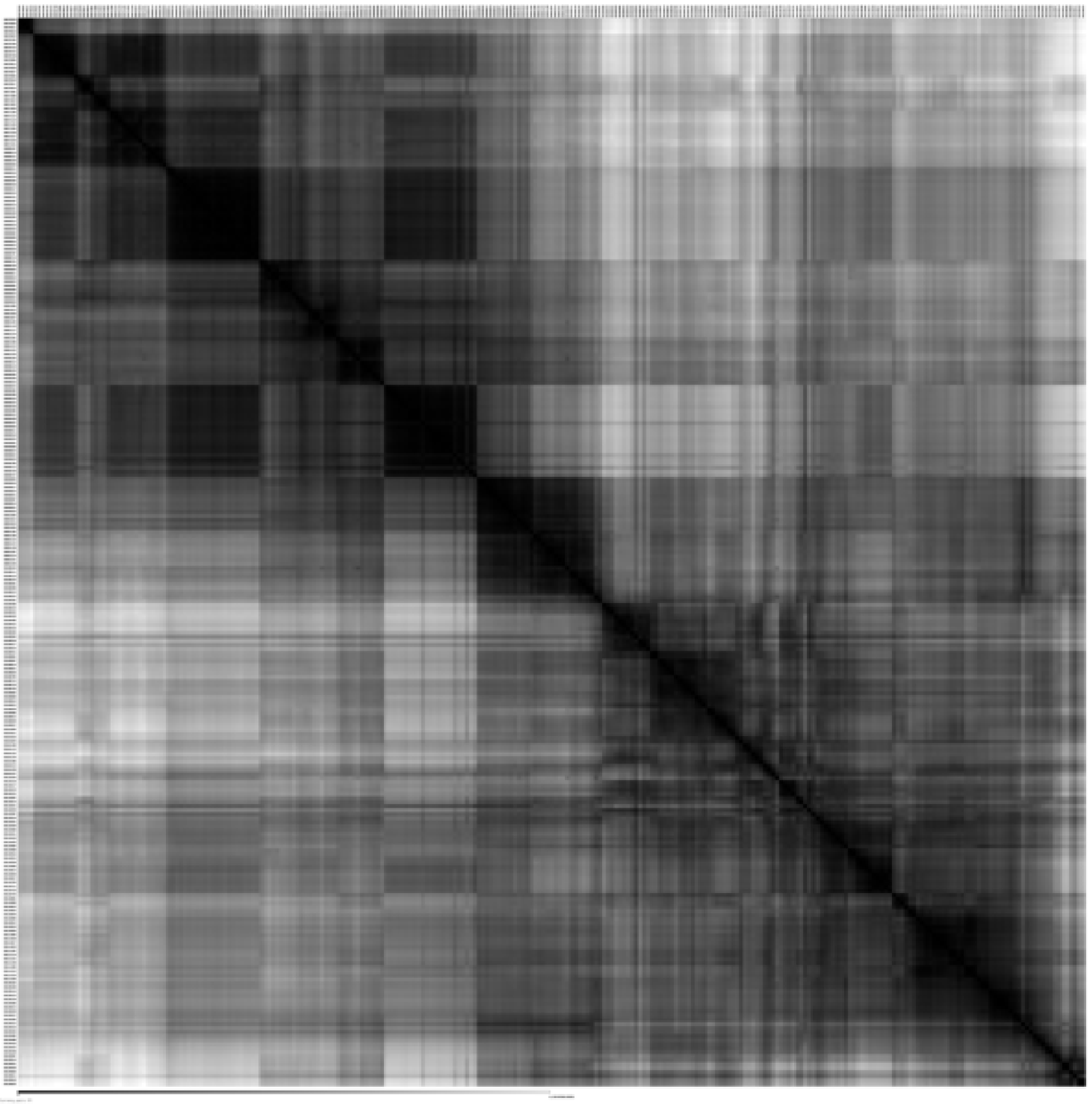}(a)
\includegraphics[scale=0.34]{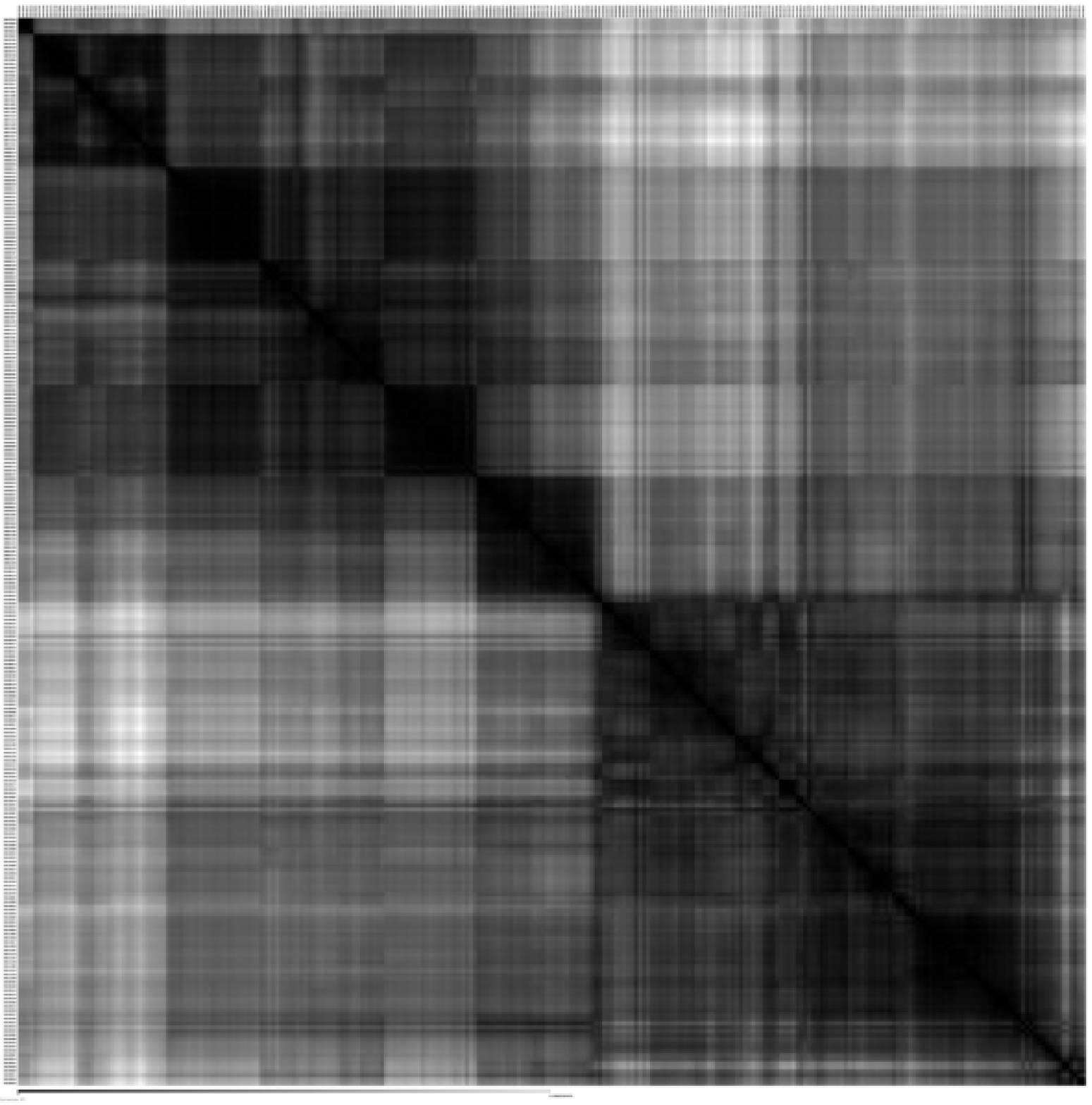}(b)
\caption{The similarities of market states between observation periods
computed from activities of currency pairs (a) and
 of currencies (b) 
 computed from quotations for the period from June 2007 to June
 2012. The black pixel represents a similar relation, and the white
 pixel represents a dissimilar relation.} 
\label{fig:DP}
\end{figure}

\begin{figure}[hbt]
\centering
\includegraphics[scale=0.34]{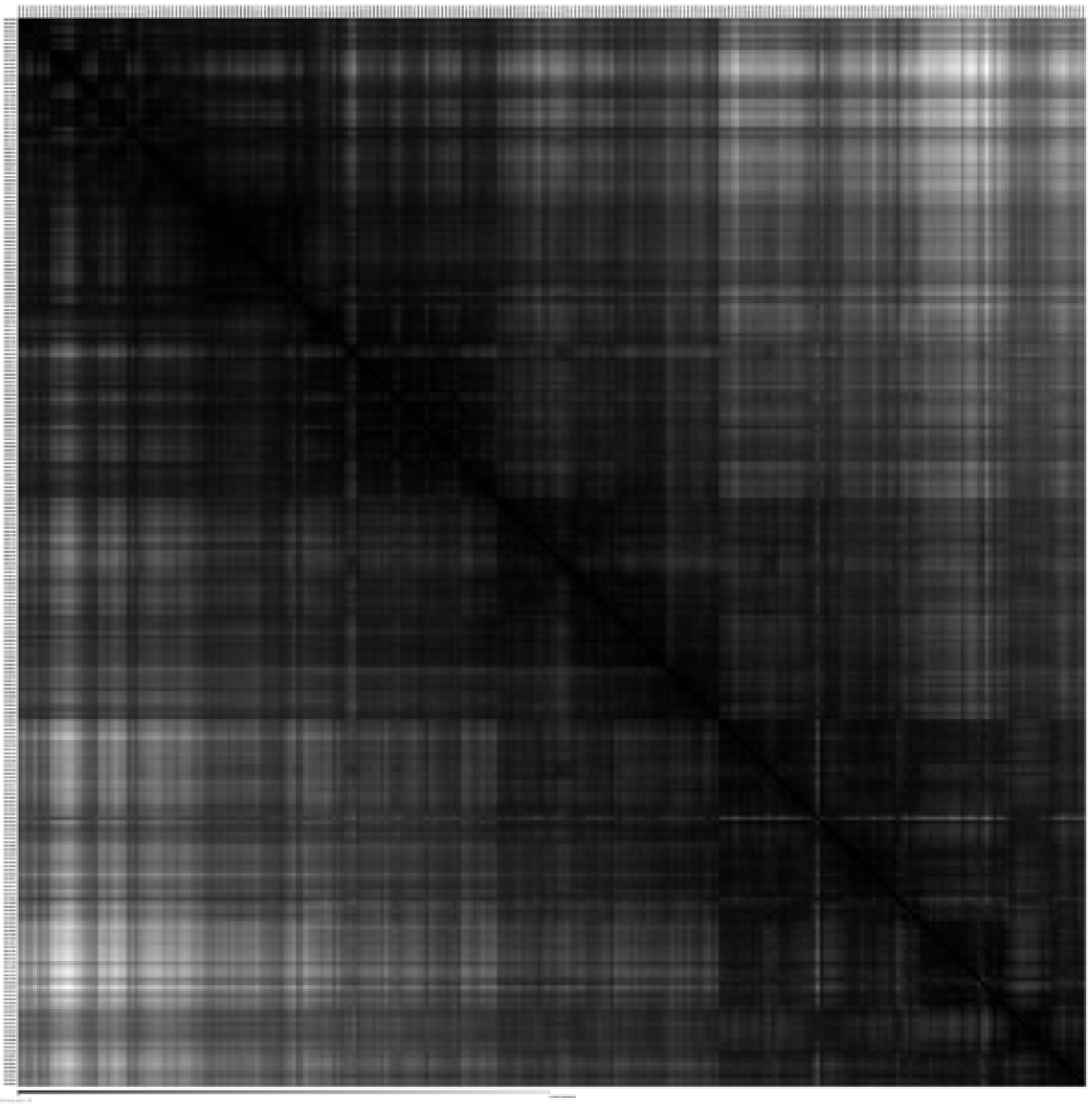}(a)
\includegraphics[scale=0.34]{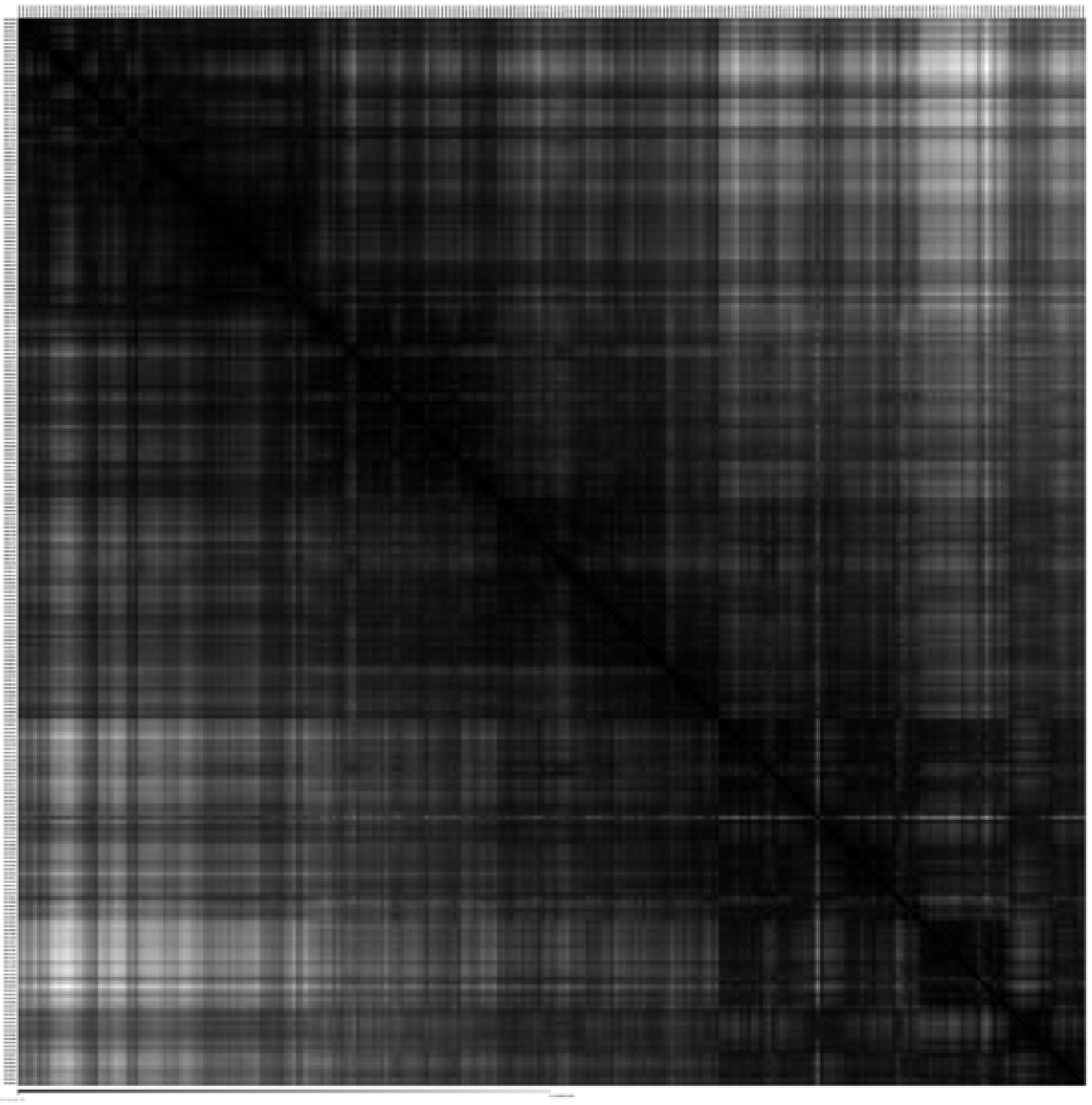}(b)
\caption{The similarities of market states between observation periods
computed from activities of currency pairs (a) and
 of currencies (b) computed from transactions for the period from
 June 2007 to June 2012. The black pixel represents a similar
 relation, and the white pixel represents a dissimilar relation.} 
\label{fig:DD}
\end{figure}

Figs. \ref{fig:DP} and \ref{fig:DD} show similarities
calculated from $A_{ij}(S)$ of quotations (a), $K_i(S)$ of quotations (b)
and similarities obtained from $A_{ij}(S)$ of transactions (a), and
$K_i(S)$ of transactions (b), respectively.


From these figures, there are regimes where shares are
equivalent. Macroeconomic situations may correspond to these regimes. 
For example, there are three regimes for the period from January 2008 to
June 2009 as shown in Fig. \ref{fig:DP}. These regimes are related to
the last global financial crisis. Furthermore, we can confirm the Euro
shock which happened March to May 2010. We can capture both the past and
present circumstances of the international economy from the high-resolution
data, and may predict or infer the future affair. It is also found that there
is an extreme regime in 14 to 21 March, 2011. This event was observed with the
sudden increase of JPY in relation to other currencies (specifically USD and
EUR). This chaos was triggered by the devastation caused by the Great East
Earthquake of Japan and the subsequent tsunami.

\subsection{Domestic hotel booking data}
Recent technological developments enable us to purchase various kinds of
items and services via e-commerce systems. The emergence of Internet
applications has had an unprecedented impact on our ability to
purchase goods and services. From the data available about items and
services on e-commerce platforms, we may expect that utilities of agents in
socio-economic systems are directly estimated. 

It is found that it is becoming more popular to make reservations of
hotels via the Internet. When we use a hotel booking site, we notice
that we sometimes find preferable room opportunities. In other words,
hotel vacancies seem to be random. We further know that both the date
and place of stay are important factors to determine the availability of
room opportunities. Hence, room availability depends on the calendar
(weekdays, weekends, and holidays) and regions. 

This availability of hotel rooms may indicate future migration
trends of travellers. Therefore, it is worth considering the 
accumulation of comprehensive data of hotel availability in order to
detect inter-migration within countries. 

Here, we give a brief explanation of a method for collecting data on hotel
availability. In this study, we used a web API in order to collect the
data from a Japanese hotel booking site named {\it Jalan}. The data is
provided by the Jalan web service~\cite{jalan}. Jalan is one of
the most popular hotel reservation services in Japan. The API is an
interface code set designed for the purpose of simplifying
the development of application programs.

The Jalan web service provides interfaces for both hotel managers
and customers. The mechanism of Jalan is as follows: Hotel
managers can enter information on room vacancies at their
hotels via an web interface. The consumers can book rooms from
available opportunities via the Jalan web site. Third parties can
even built their web services with Jalan data by using the web API.

We collected all available opportunities which appeared on
Jalan web regarding one-night room vacancies for two adults. 
The data were sampled from the Jalan web service daily. The data on room
opportunities collected through the Jalan web API are stored as
comma separated value (csv) files.

In the data set, there exist over 100,000 room opportunities at over
14,000 hotels. In Table \ref{tab:data}, we show contents included in the
data set. Each plan contains sampled date, stay date, regional
sequential number, hotel identification number, hotel name, postal
address, URL of the hotel website, geographical position, plan name,
and rate.

Since the data contains regional information, it is possible for us to
analyse regional dependence of hotel rates. Throughout the
investigation, we regard the number of recorded opportunities (plan) as
a proxy variable of the number of available room stocks. 

\begin{table}[h]
\caption{The data format of room opportunities.}
\label{tab:data}
\centering
\begin{tabular}{l}
\hline
\hline
Date of collection \\
Date of stay \\
Hotel identification number \\
Hotel name \\
Hotel name (kana characters) \\
Postal code \\
Address \\
URL \\
Latitude \\
Longitude \\
Opportunity name \\
Meal availability \\
The latest best rate \\
Rate per night \\
\hline
\hline
\end{tabular}
\end{table}

\begin{figure}[hbt]
\centering
\includegraphics[scale=0.3]{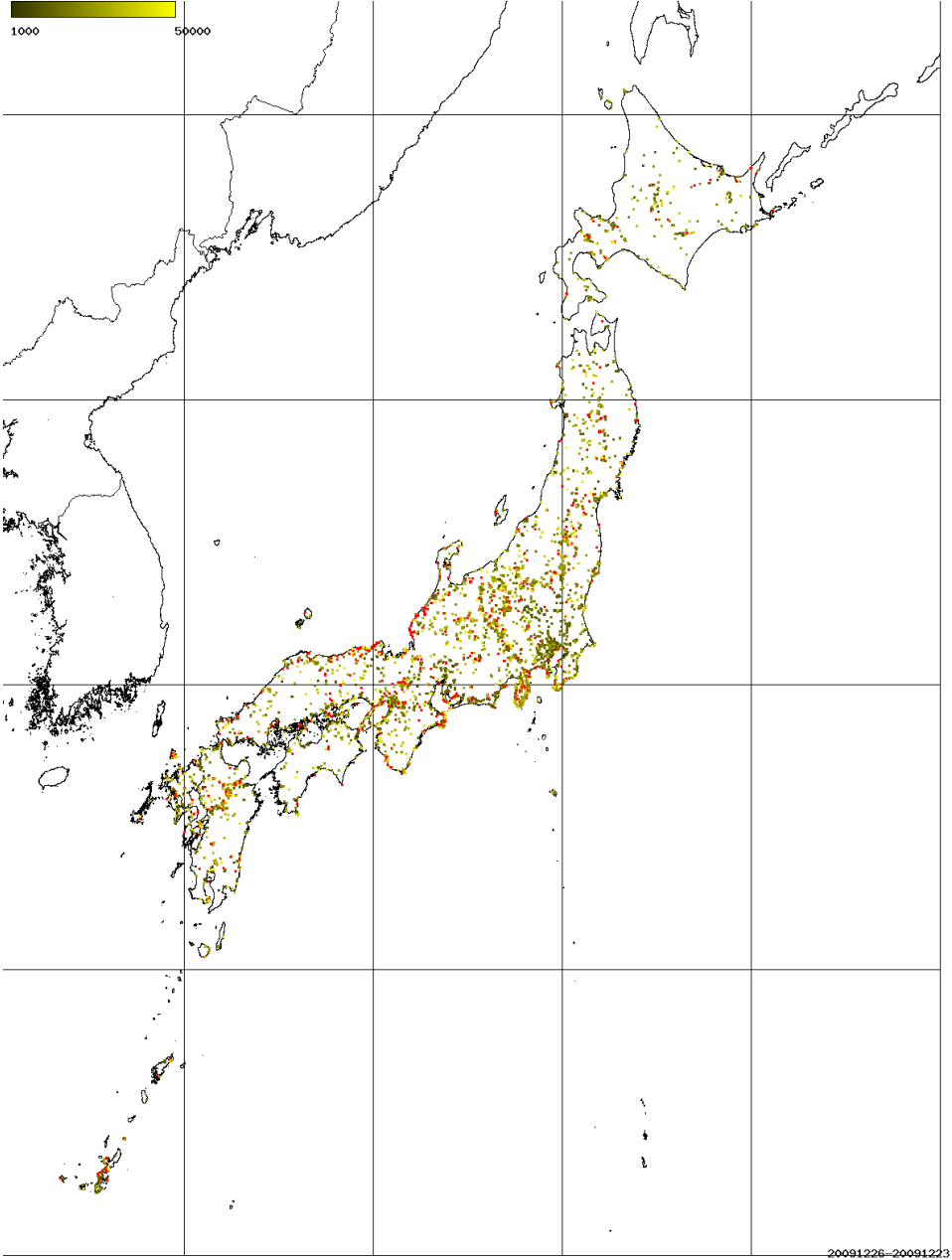}
\caption{The regional dependence of room prices of Japanese
 available on 26 December, 2009, as of 23 December, 2009.}
\label{fig:map}
\end{figure}

For this analysis, we used the data for the period from 24 December, 2009 to
8 May, 2011. The data is missing from 14 to 30 March, 
2011 because the the web service was not available due to the Great East 
Earthquake. Fig. \ref{fig:map} shows an example of distributions
and representative rates. An example of rates distributions under the
condition that two adults can stay at the hotel for one night at 23
December, 2009. This data have been sampled on 25 December, 2009. 
The yellow to red filled squares represent hotel plans costing ranging
from 50,000 to JPY 1,000 JPY per night. The red filled squares represent
hotel plans costing over 50,000 JPY per night. We found that there was a
strong dependence of vacancies on places. Specifically, we find that
many hotels are located around several centralised cities such as Tokyo,
Osaka, Nagoya, Fukuoka, and so on. 

\begin{figure}[hbt]
\centering
\includegraphics[scale=1.0]{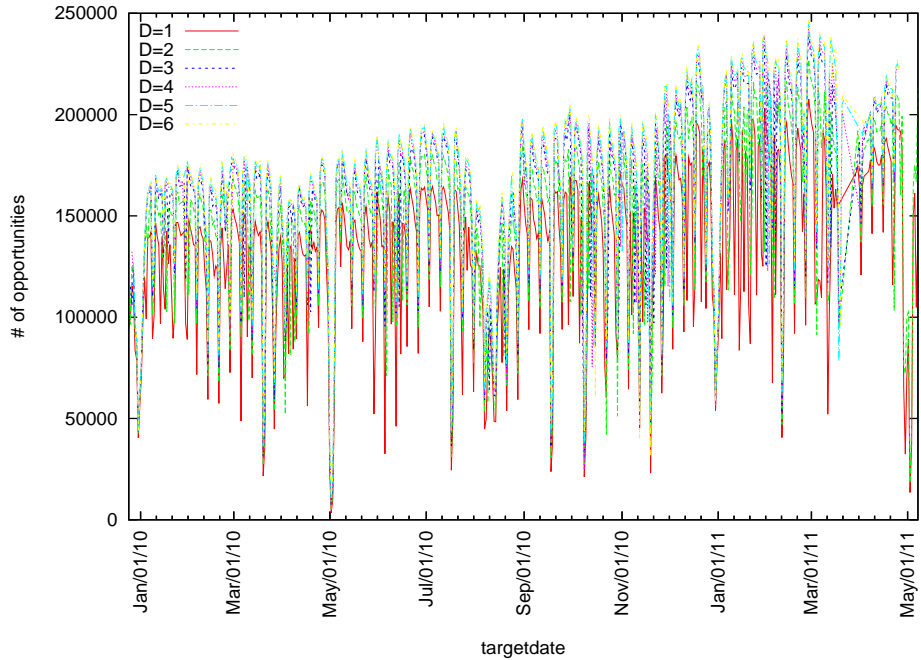}
\caption{The number of room plans for one-night stay
by two adults during the period between 24 December, 2009 to 8 May, 2011.}
\label{fig:number}
\end{figure}

The number of one-night, twin-share room plans was
counted from the recorded csv files throughout the whole sampled 
period. Fig. \ref{fig:number} shows the daily number of room
opportunities with different durations $D$, which is defined as a
difference between stay date and booking date. From this graph, we found
three facts:

\begin{description}
\item{(1)} The number of room opportunities fluctuates weekly.
\item{(2)} There is a strong dependence of the number of available
opportunities on the Japanese calendar. Namely, Saturdays and holidays
drove reservation activities of consumers. For example, during the New
Year holidays (around 12/30-1/1) and holidays in the spring season
(around 3/20), the time series of the numbers show big drops. 
\item{(3)} The number eventually increases as the date of stay
reaches. Specifically, it is observed that the number of opportunities
drastically decreases two days before the date of stay.  
\end{description} 

\begin{figure}[hbt]
\centering
\includegraphics[scale=1.0]{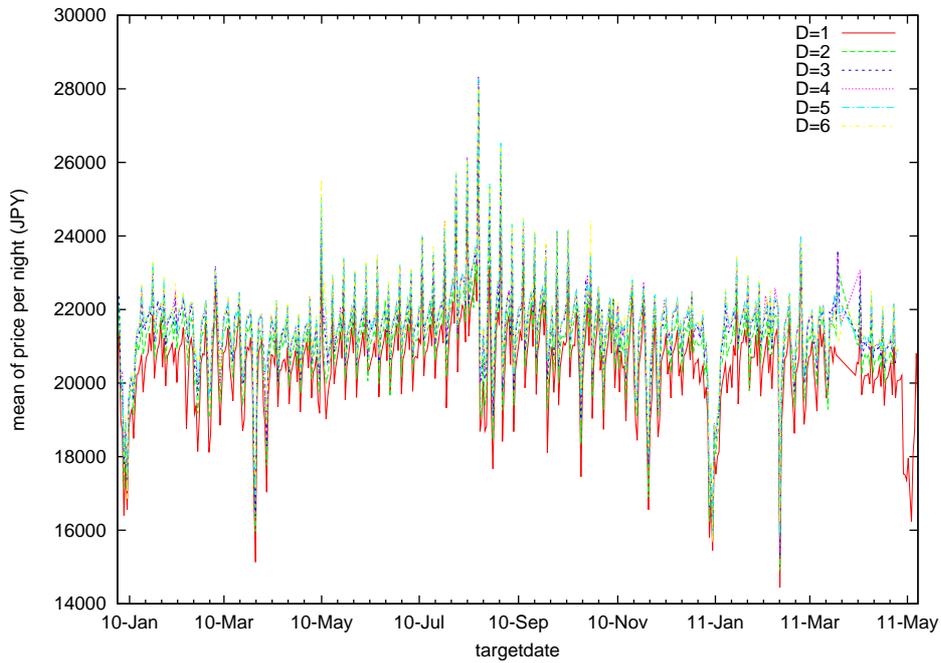}
\caption{Average prices for one-night, twin-share room plans
throughout Japan for the period from 24 December, 2009 to 8 May, 2011.}
\label{fig:mean}
\end{figure}

Furthermore, Fig. \ref{fig:mean} shows dependence of average rates all over the Japan
on calendar dates with different durations. During the New Year holidays in
2010, the average rates rapidly decreased.
Meanwhile, on the spring holidays in 2010, the average rates rapidly
increased. This difference seems to arise from the difference of
consumers' motivation structure and preference on price levels between
these holiday seasons.  

\begin{figure}[hbt]
\centering
\includegraphics[scale=1.0]{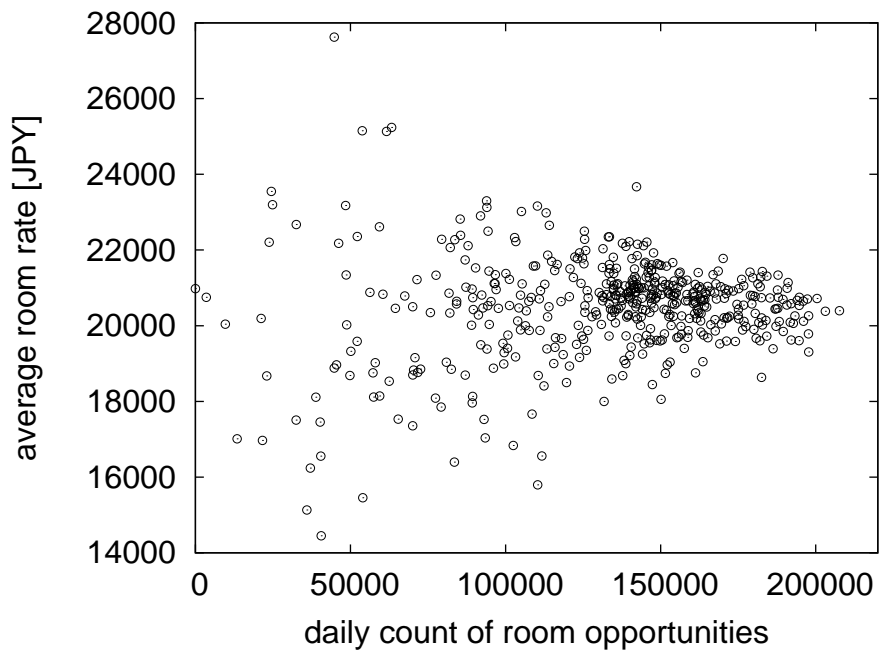}
\caption{Scatter plots between the daily number of room opportunities
 and mean room rates across Japan for the period from 24 December,
 2009 to 10 June, 2011.}
\label{fig:count-mean}
\end{figure}

Figure \ref{fig:count-mean} shows scatter plots between the daily number
of room opportunities and average of room rates. The high-demand dates
exhibit larger variations of the average rate than low-demand
dates. The preferable price level of consumers has a high variability on
high-demand dates.

\subsection{Impact of natural disasters (Great East Japan Earthquake on
11 March, 2011)}

Since people are also products of nature, the physical
effects of the natural environment on our society are
remarkable. Specifically, natural disasters often affect our societies
significantly. Therefore, we need to understand the subsequent impact of
natural disasters on human behaviour, from both economical and social
perspectives. 

The first Great East Japan Earthquake hit at 14:46 on 11 March, 2011 in
Japanese local time (05:46 in UTC). Within 20 minutes,
huge tsunamis had devastated cities along Japan's northeastern
coastline. In addition to wide-spread physical destruction,
social infrastructures also suffered extensive damaged. It is
important for us to understand its subsequent impact on our
socio-economic activities. 

In order to estimate both economic and social damages in three Tohoku
prefectures (Iwate, Miyagi, and Fukushima), we focus on the number of
available hotels in each district before and after the Great East Japan
Earthquakes and Tsunami. We selected 21 districts in three prefectures,
as shown in Table I and two periods, one before and one after the
disaster. The data on accommodation in this area cover about 31\% of the
potential accommodation. Therefore, we have to estimate the states
that were not sampled from these sampled booking data. 

If we assume that the accommodations in the data are sampled from uncensored
data in a homogeneous way, then the relative frequency of the available
accommodation from censored data can approximate the true value,
computed from uncensored data. The data on accommodations in this area cover
about 31\% of the potential accommodation. Therefore, we have to estimate the
uncensored states from these censored booking data. 

If we assume that accommodation in the data are sampled from
uncensored data in a homogeneous way, then a relative frequency of the
available accommodations from censored data can approximate 
the true value that would be computed from uncensored data.

In order to conduct a qualitative study, let $x_i(t,s) \quad (i=1,\ldots,K;
t=1,\ldots, T)$ be the number of available hotels in district $i$ at day
$t$ in period $s$. Then a relative frequency at district $i$ can be
calculated as,
\begin{equation}
p_i(s) = \frac{\sum_{t=1}^T x_i(t,s)}{\sum_{i=1}^K \sum_{t=1}^T x_i(t,s)}.
\end{equation}
Let us consider a ratio of the relative frequencies after and before a
specific event,
\begin{equation}
q_i(a|b) = p_i(a)/p_i(b),
\label{eq:ratio}
\end{equation}
where $p_i(a)$ and $p_i(b)$ represent the relative frequencies after and
before the event, respectively. Obviously, Eq. \ref{eq:ratio} can
be rewritten as:
\begin{equation}
q_i(a|b) = \frac{n_i(a)}{n_i(b)}/\frac{N(a)}{N(b)},
\label{eq:change-of-ratio}
\end{equation}
by using $n_i(s)$ and $N(s)$, which are the number of available hotels
at district $i$ within the period $s$ and the total number during that
period. Since $N(a)/N(b)$ is independent of $i$, $q_i(a|b)$ should be
proportional to a ratio of the number of hotels after and before the event.

Table \ref{tab:district} shows $q_i(a|b)$, where the term $b$ represents
May 2010 (before the disaster), and the term $a$ May 2011 (after the
disaster), respectively. Since the value of $q_i(a|b)$ is related to
damage to hotels in the district $i$, $q_i(a|b)<1$ implies that
available hotels decreased after the earthquake 
at $i$ relative to the total number of hotels. Similarly
$q_i(a|b)>1$ means that they maintained at $i$. 

We may assume that the decrease of $q_i(a|b)$ at district $i$ results
from both a decrease of supply and an increase of demand. The decrease of
supply is caused in this case by the physical destructions of
infrastructure. The increase of demand comes from behaviour of individuals like
refugees, workers, volunteers, and civic groups.

The regional dependence of supply can be estimated from the number of
destroyed houses in each district. To do so, we calculate the numbers of
both completely-destroyed and partially-destroyed houses at each
district from the data downloaded from a website of the National Research
Institute for Earth Science and Disaster Prevention~\cite{NDIS}. 
The numbers are calculated by summing the number of destroyed houses 
in the towns or cities included in each district. Table \ref{tab:district} shows
the numbers of destroyed houses. In this histogram it is shown that
damaged houses were concentrated in the maritime areas of these prefectures.

\begin{table}[h]
\caption{The ratio of the number of available hotels during the period from
1st to 31st May 2011 to that during the period from 1st to 31st May 2010
(after and before the Great East Japan Earthquake), the number of
 both completely destroyed houses and partially destroyed houses, as
 confirmed at the end of September 2011 and the number of evacuees, as
 confirmed at 1st May 2011.}
\label{tab:district}
\centering
\begin{tabular}{lllllll}
\hline
\textit{prefecture} & \textit{district} & \textit{ratio} & \textit{complete collapse} & \textit{partial collapse} & \textit{evacuees} \\
\hline
Iwate & Shizukuishi & 1.970 & 0 & 0 & 372 \\
      & Morioka & 1.834 & 0 & 4 & 366 \\
      & Appi,Hachimantai,Ninohe & 2.250 & 3 & 0 & 0 \\
      & Hanamaki,Kitakami,Tohno & 1.350 & 27 & 364 & 853 \\
      & SanrikuKaigan & 0.481 & 18,098 & 2,166 & 12,896 \\
      & Oushu,Hiraizumi,Ichinoseki & 0.374 & 83 & 533 & 338 \\
\hline
Miyagi & Sendai & 0.550 & 21,789 & 37,522 & 3,608 \\
       & Matsushima,Shiogama & 0.345 & 7,895 & 12,581 & 5,115 \\
       & Ishinomaki,Kesennuma & 0.0 & 33,661 & 6,083 & 23,840 \\
       & Naruko,Osaki & 1.484 & 486 & 1,577 & 929 \\
       & Kurihara,Tome & 1.404 & 224 & 1,105 & 1,049 \\
       & Shiroishi,Zao & 1.608 & 2,522 & 1,644 & 1,612 \\
\hline
Fukushima & Fukushima,Nihonmatsu & 0.665 & 168 & 1,898 & 1,321 \\
          & Soma & 0.038 & 6,279 & 1,618 & 1,969 \\
          & Urabandai,BandaiKogen & 1.134 & 0 & 0 & 2 \\
          & Inawashiro,Omotebandai & 1.009 & 10 & 12 & 303 \\
          & Aizu & 1.352 & 4 & 27 & 266 \\
          & Minamiaizu & 1.768 & 0 & 0 & 14 \\
          & Koriyama & 0.604 & 2,596 & 12,185 & 2,489 \\
          & Shirakawa & 1.915 & 135 & 1,820 & 418 \\
          & Iwaki,Futaba & 0.195 & 6,550 & 17,614 & 2,115\\
\hline
\end{tabular}
\end{table}

\begin{table}[h]
\caption{The number of evacuees of the Great East Japan Earthquake at
 three prefectures (Iwate, Miyagi, and Fukushima). The data were officially
 announced by the Japanese Cabinet Office on 3rd June 2011.}
\label{tab:victims}
\centering
\begin{tabular}{lllll}
\hline
prefecture & A: public places & B: hotels & C: others & A+B+C \\
\hline
Aomori & 0 & 78 & 777 & 855 \\
Iwate & 9,039 & 2,007 & 14,701 & 25,747 \\
Miyagi & 23,454 & 2,035 & - & 25,489 \\
Akita & 128 & 619 & 909 & 1,656 \\
Yamagata & 305 & 779 & 2,366 & 3,450 \\
Fukushima & 6,105 & 17,874 & - & 23,979 \\
\hline
\end{tabular}
\end{table}

We can confirm that house damage was serious in
Sanrikukaigan, Sendai, Matsushima, Shiogama, Ishinomaki, Kesennuma,
Soma, Koriyama, Iwaki, and Futaba. The greatest number of
completely-destroyed houses is 33,661 in Ishinomaki and Kesennuma. The
second is 21,789 in Sendai. The third is 18,098 in Sanrikukaigan. The
greatest number of partially-destroyed houses is 37,522 in
Sendai. The second is 17,614 in Iwaki and Futaba. The third is 12,185 in
Koriyama. 

In fact, in places where the ratio $q_i(a|b)$ is greater than 1, 
the number of destroyed houses is not significant, as shown in
Figure \ref{tab:district}. We confirmed that the ratio $q_i(a|b)$ may measure
the degree of damage to economic activity in the travel industry.
However, it is not confirmed that there was significant physical 
damage to houses in Oushu, Hiraizumi, Ichinoseki, Fukushima, and
Nihonmatsu, even having a ratio less than 1. It may be thought that
hotels in Oushu, Hiraizumi, and Ichinoseki were used by workers and
evacuated victims of the disaster. Decreases of available hotels in
Fukushima and Nihonmatsu may be related to accidents in Fukushima
Daiich nuclear power plant. The number of victims evacuated from
the disaster in each prefecture, according to an official 
announcement by the Japanese Cabinet Office on 3 June 2011,
is shown in Table \ref{tab:victims}. In the case of Fukushima
prefecture, 17,874 people were evacuated to hotels at that time.

\subsection{International flight booking data}
AB-ROAD~\cite{abroad} is a Japanese Internet travel booking
site. About 14,000 flights are available on this site every
day. This booking site serves as a web API for both travel agencies and
customers. On the one hand, travel agencies can register their flight
opportunities on the site via the Internet. On the other hand, consumers
can search for and book flights that they want to purchase from all the
registered flights via the website. Third parties can even build web
services with the data provided by the web API. 

I collected information regarding available flight tickets using the
AB-ROAD web service every day and stored it as comma-separated (CSV)
files. This data set contains the flight tickets that a person would be
able to use to depart from one of the airports in Japan. Each flight
also contains the date when the data was sampled, departure date,
departure airport, arrival airport, type of class (economy, business,
and first), name of air carrier, and price (the fuel surcharge and tax
are excluded). The data period is from 29 July, 2010 to 14 December,
2011. For technical reasons, data on several dates is missing (8
November, 2010, 10 April, 2011, from 14 to 25 April, 2011).  

We use data of opportunities where flights would depart from some airports 
in Japan 4 weeks later. The supply-demand situation of air travels may
be related to the number of opportunities. The price of
flight tickets between departure and arrival airports is also
dependent on their distance. Fundamentally, I investigates current
situations of international air flights departing from Japanese airports.
In the dataset, there exist about 14,000 kinds of flight
opportunities for about 78 airline companies~\footnote{
The included airline companies are listed as follows: Jetstar Asia
Airways (3K), Cebu Air (5J), Jeju Air (7C), Gill Airways (9C), Jet
Airways (9W), American Airline (AA), Air Canada (AC), Mandarin
Airlines (AE), Air France (AF), Air India (AI), Aeromexico (AM),
Finnair (AY), Alitalia (AZ), British Airways (BA), Eva Air (BR), Air
Busan (BX), Air China (CA), China Airlines (CI), Continental
Airlines (CO), Cathay Pacific Airways (CX), China Southern
Airlines (CZ), Delta Air Lines (DL), Emirates (EK), Etihad Airways
(EY), Shanghai Airlines (FM), Garuda Indonesia (GA), Hawaiian
Airlines (HA), Hong Kong Airlines (HX), Uzbekistan Airways (HY),
Business Air (II), Iran Air (IR), Air Inter (IT), Japan Airlines (JL),
JALways (JO), Jetstar Airways (JQ), Korean Air (KE), KLM-Royal
Dutch Airlines (KL), Kenya Airways (KQ), Lufthansa German
Airlines (LH), Crossair (LX), Air Madagascar (MD), Xiamen
Airlines (MF), Malaysia Airline System Berhad (MH), SilkAir (MI),
EgyptAir (MS), China Eastern Airlines (MU), All Nippon Airways
(NH), Northwest Airlines (NW), Air Macau (NX), Air New Zealand
(NZ), MIAT Mongolian Airlines (OM), Austrian Airlines (OS),
Asiana Airlines (OZ), Pakistan International Airlines (PK), Philippine
Airlines (PR), Air Niugini (PX), Qantas Airways (QF), Qatar
Airways (QR), Cargolux (S1), South African Airways (SA), Air
Caledonie International (SB), Shandong Airlines (SC), Scandinavian
Airlines (SK), Brussels Airlines (SN), Singapore Airlines (SQ),
Aeroflot (SU), Thai Airways (TG), Turkish Airlines (TK), Air Tahiti
Nui (TN), United Airlines (UA), Air Lanka (UL), Transaero Airlines
(UN), Hong Kong Express Airways (UO), Vietnam Airlines (VN),
Virgin Atlantic (VS), Vladivostok Air (XF), Arcus Air (ZE) and
Shenzhen Air (ZH).} every day.  

\begin{figure}
\centering
\includegraphics[scale=0.9]{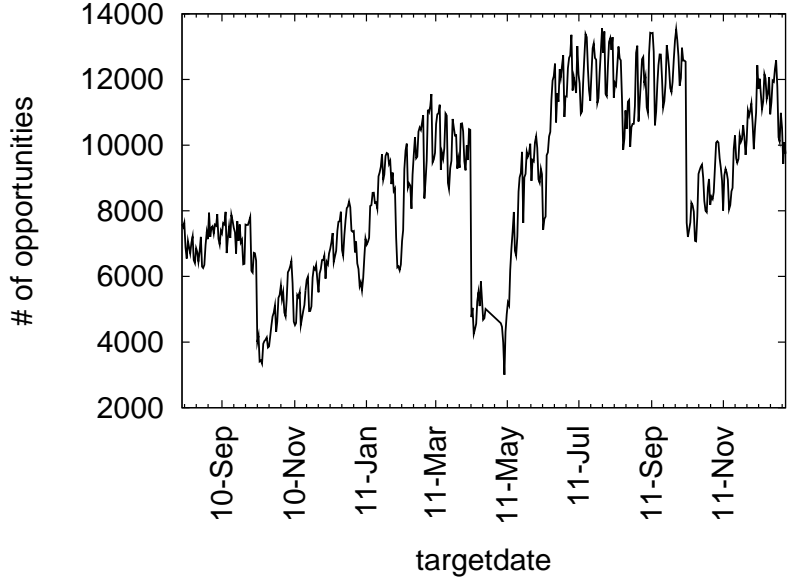}
\caption{The daily number of flight opportunities from 29 July, 2010 to
 24 December, 2011.}
\label{fig:flightnumber}
\end{figure}

The total number of available flights from a city in Japan
to a city in a foreign country was totalled from the data throughout the
entire sampled period. Figure \ref{fig:flightnumber} shows the total
number of available flights per day. From observing this data,
we made three observations:

\begin{description}
\item{(1)} There exists weekly seasonality for the total number of available
flight tickets. The demand of flight tickets is higher on Sundays and
Mondays than on other days.
\item{(2)} The number of flight tickets strongly depends on the Japanese
calendar. Namely, summer holidays influence the reservation activities
of consumers. For example, during the Golden Week holidays (from 1 to 5 May,
2011) and the holidays in the spring season (around 20 March, 2011),
total availability shows steep decreases. 
\item{(3)} Since several airline companies update their flight schedule every
April and October, ticket availability drastically drops at that time. 
\end{description}

Recently, the airport network has been studied by several
researchers~\cite{Guimera,Guida,Bagler}. 
According to the study by Guimer\`a and Amaral~\cite{Guimera}, the
world-wide airport network has properties of a small-world network. The
degree and betweenness centrality distributions exhibit power-law
decay. In fact, the most connected cities (largest degree) are typically
not the most central cities (largest betweenness centrality). Airports
with high betweenness tend to play a more important role in keeping
networks connected than those with high degree, because a passenger can
travel from a departure airport to a destination with a short
path. The geodesic distance between departure and arrival
airports may give a good approximation of the actual flight distance of
passengers.

\begin{figure}[hbt]
\centering
\includegraphics[scale=0.5]{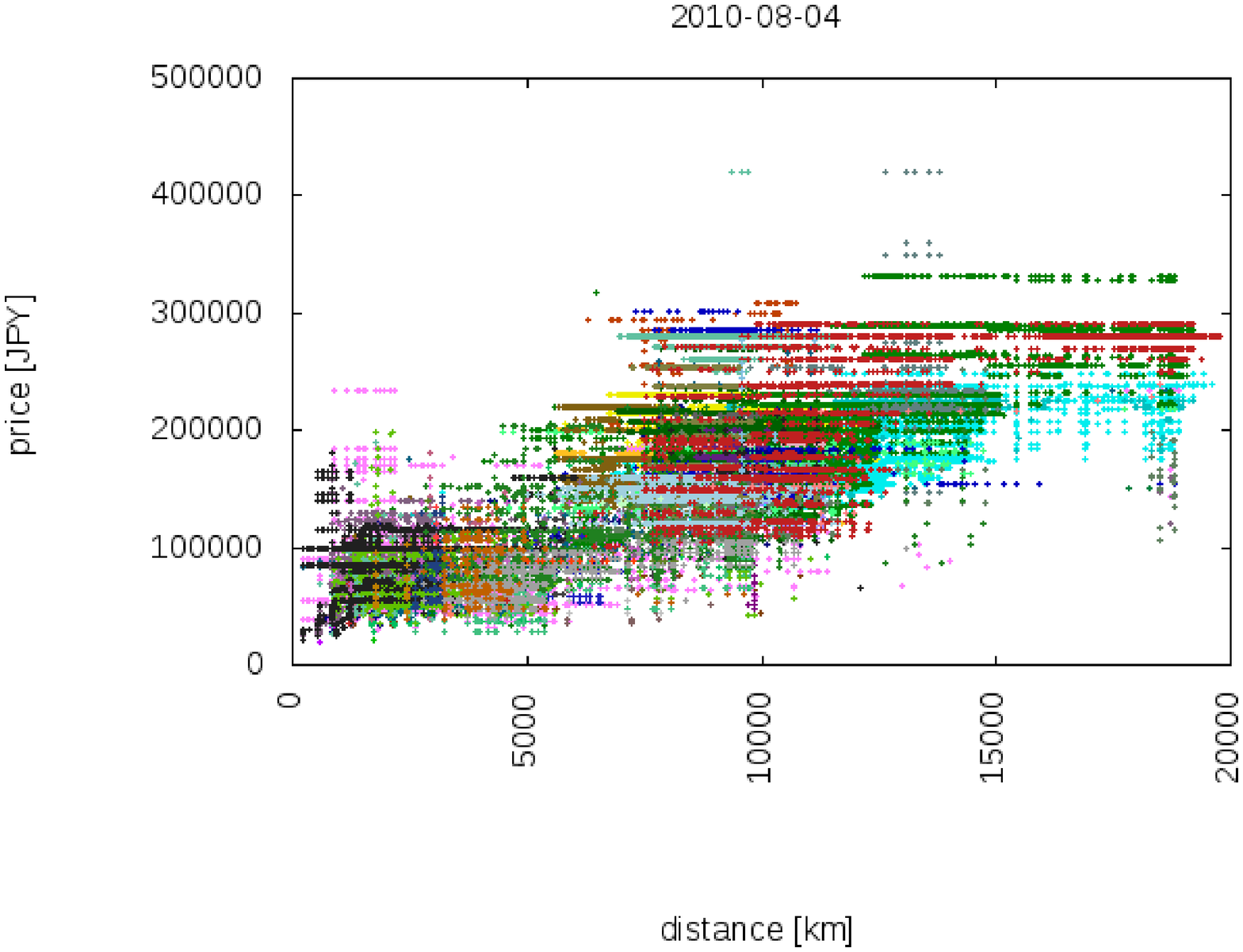}(a)
\includegraphics[scale=0.5]{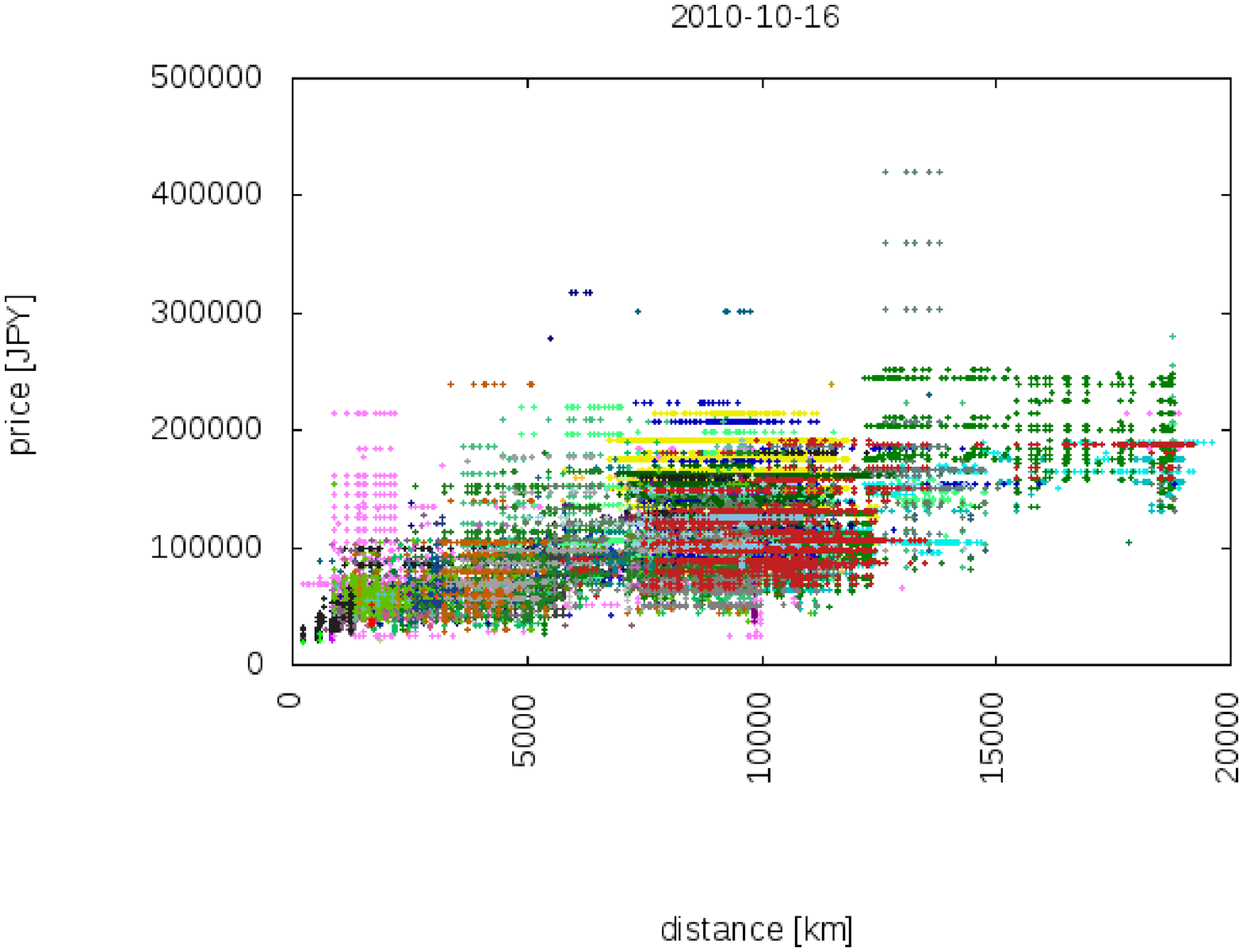}(b)
\caption{Relationship between the price of economy-class flights
 and geodesic distance on 4 August, 2010 (a) and on 16
 October, 2010 (b).}
\label{fig:flight}
\end{figure}

The geodesic distance is measured by Vincenty's formulae. Let $\phi_s$,
$\lambda_s$, $\phi_f$ and $\lambda_f$ be the geographical latitude and
longitude of two points, respectively, and $\Delta \lambda =
\lambda_s-\lambda_f$. Under the assumption that the earth is a sphere,
the distance $D$ is given by,
\begin{equation}
D = r\tan^{-1}\Bigl(\frac{\sqrt{(\cos\phi_s \sin \Delta \lambda)^2 +
 (\cos\phi_s \sin\phi_f - \sin\phi_s \cos\phi_f \cos\Delta
 \lambda)^2}}{\sin\phi_s\sin\phi_f + \cos\phi_s\cos\phi_f \cos\Delta \lambda}\Bigr),
\end{equation}
where $r$ represents the earth's radius ($r = 6371.2$ [km]). It is
possible to analyse the geodesic dependence of ticket prices with 
this data. Figure \ref{fig:flight} shows the relation between the price of
economy-class flight tickets and the geodesic distance from the
departure airport to the destination. The distance of each flight ticket
is computed from the geographical latitude and longitude of the
departure and arrival airports using Equation (1). Fig. \ref{fig:flight}
represents the relationship on 4 August, 2010 (high demand season) and
on 16 October, 2010 (low demand season). Short-distance corresponds to
flights to Asian cities (1,000 km to 3,000 km), middle-distance to
cities in Europe and North America (8,000 km to 10,000 km), and
long-distance to cities in South America (15,000 km to 20,000 km). During
high demand season, it is found that various kinds of flights appear for
both short-distance and long-distance flights, but, during low demand
season, there are few long-distance flights. 

\section{Discussion}
\label{sec:discussion}
It was shown from looking at these exemplar studies that 
there are two kinds of axes with which to best represent our society.
The studies of the Japanese stock exchange
market and the foreign exchange market show that different stocks or
currencies are quoted or traded concurrently, and that they are mutually
related. Namely, time is one of important key factors to link data
observed in different places. Investigation of both the hotel and
international flights showed that geographical information is also a key factor
to link data from different layers. 

Since events in the world happen in massively concurrent, time and
locations are the most important elements to link results obtained from
different data sources. Furthermore, we may synthesise data from
different data sources with a time and location, and construct a
large-scale database.

I also want to say that these studies are of
econoinformatics. Econoinformatics  
includes several research topics such as mathematical concepts,
algorithms, computer systems, databases, modelling, and so
forth. Mathematical concepts may help us to develop automated algorithms
to extract knowledge on socio-economic systems. The algorithm and
computation systems help us to collect, store and analyse large amounts
of data. It is also crucial to construct synthesised data sets from
uncensored data obtained from different data sources. Furthermore,
mathematical models will become a guide for us to understand observed
systems connecting physical mechanisms with observations. 

\begin{figure}[!hbt]
\centering
\includegraphics[scale=0.5]{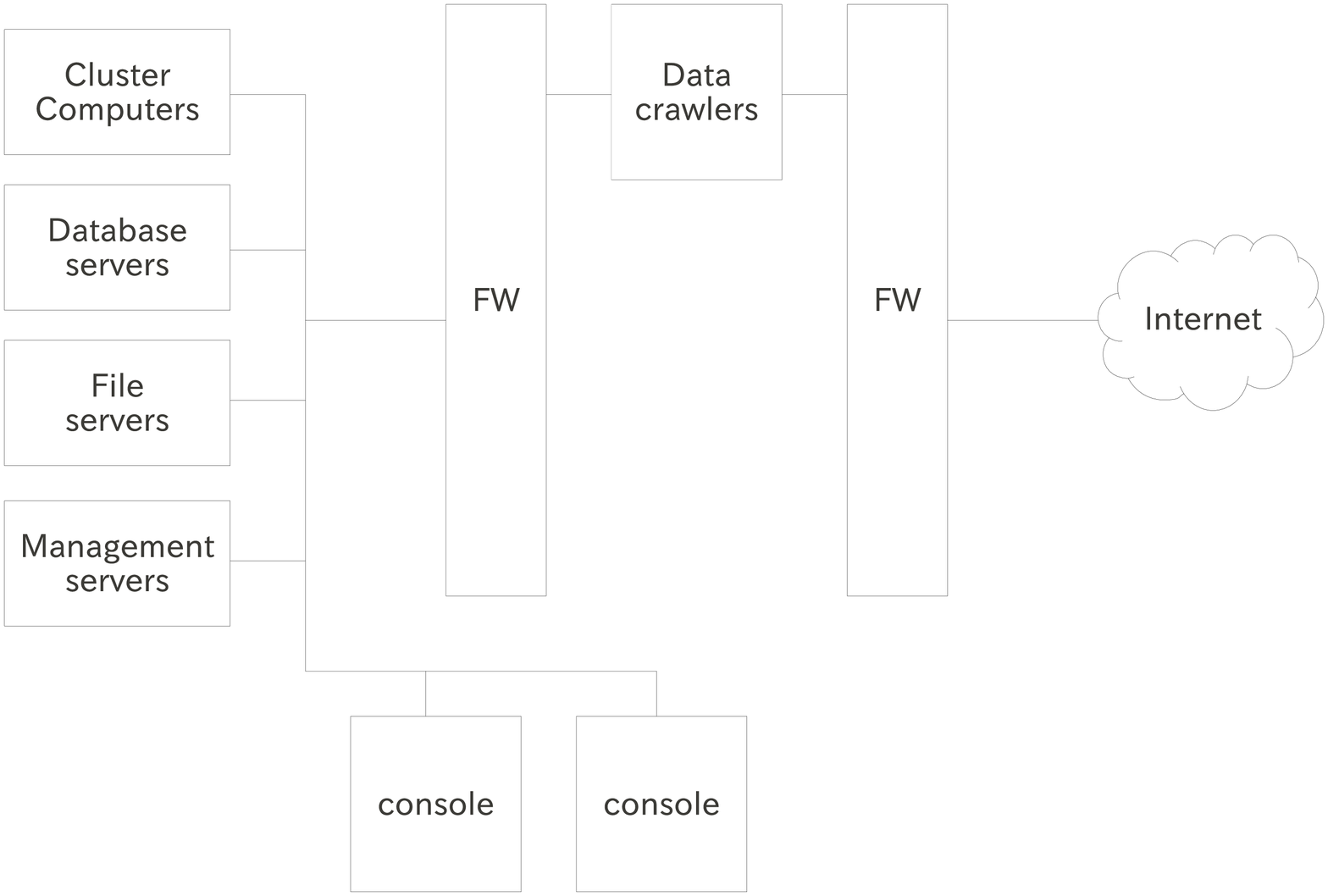}
\caption{A conceptual illustration of the research platform.}
\label{fig:system}
\end{figure}

Figure \ref{fig:system} shows a conceptual illustration of a research
platform. This consists of data crawlers, cluster computers, database
servers, management servers, file servers, and consoles. The data
crawlers collect socio-economic data from e-commerce platforms in the
internet. The data is stored in the file servers. After verifying the
data, the data is synthesised in the database servers. To construct
the synthesised database, the data is stored with the time and/or
locations. The management servers handle this procedure of collecting,
verifying, and storing data automatically. In the cluster computers, the
socio-economic data is computed and analysed. Researchers use this system
from the consoles.

\section{Conclusion}
\label{sec:conclusion}
The concept of complexity was discussed from several points of
view. Complexity is determined by the relative degrees of freedoms of 
observing systems and of observed systems. Recent exemplar
studies on data-centric socio-economic systems were reviewed and presented. 
The Japanese stock exchange, foreign exchange market, domestic hotel
bookings, and international flight bookings were analysed
from a comprehensive point of view.

Econoinformatics can be established from data-centric studies 
based on large amounts of data on socio-economic systems. Such data
can be obtained from our society. To overcome the complexity of
socio-economic data, we need to develop mathematical concepts,
algorithms, rich computer systems, 
and synthesised databases. To construct synthesised data on
socio-economic systems, spatial-temporal axes should be
employed. Various kinds of data collected from different paths may be
synthesised with time and locations. Rich synthesised databases
representing aspects of our society may help us to obtain deeper insights
into our own socio-economic systems.

\section*{Acknowledgement}
The author is thankful to Prof. Thomas Lux, Prof. Burda Zdzis\l aw,
Prof. Janusz Ho\l yst, and Prof. Dirk Helbing for stimulating discussions,
and to Ms. Youko Miura (AB-ROAD), Mr. Kotaro Sasaki (Jalan) and
Mr. Daichi Tanaka (Jalan), Mr. Hiroshi Yoshimura (Jalan) for providing
useful information on air travel and the domestic hotel industry. This work
was partially supported by the Grant-in-Aid for Young Scientists (B)
(\#23760074) by the Japanese Society of Promotion of Science (JSPS).

\appendix
\section{Proof of $0\log 0$}
\label{sec:zero-log-zero}
Let us consider
\begin{equation}
h = \lim_{x\rightarrow +0} x \log x.
\end{equation}
Putting $x = e^{-z}$ one has
\begin{equation}
h = -\lim_{x\rightarrow +0} z e^{-z}.
\end{equation}
By using the Taylor expansion of $e^{z} = \sum_{k=0}^{\infty}\frac{1}{k!}z^k$, 
one obtains
\begin{eqnarray}
\nonumber
h &=& -\lim_{z\rightarrow \infty}z e^{-z} \\
\nonumber
&=& -\lim_{z\rightarrow \infty} z / e^{z} \\
\nonumber
&=& -\lim_{z\rightarrow \infty} \frac{z}{\sum_{k=0}\frac{1}{k!}z^k} \\
\nonumber
&=& -\lim_{z\rightarrow \infty} \frac{1}{\sum_{k=0}\frac{1}{k!}z^{k+1}} \\
&=& 0.
\end{eqnarray}
Therefore, we gets
\begin{equation}
h = \lim_{x\rightarrow 0} x \log x = 0 \log 0 = 0.
\end{equation}
\end{document}